\documentclass[preprint,pre,aps,showpacs]{revtex4}
\usepackage{graphicx}
\usepackage{color}

\begin{document}

\title{Dynamical and statistical bimodality in nuclear fragmentation}

\author{S.Mallik$^1$, G. Chaudhuri$^1$ and F. Gulminelli$^2$}

\affiliation{$^1$Physics Group, Variable Energy Cyclotron Centre, 1/AF Bidhan Nagar, Kolkata700064,India\\
$^2$LPC Caen IN2P3-CNRS/EnsiCaen et Universite, Caen, France}


\begin{abstract}
 The origin of bimodal behavior in the residue distribution experimentally measured in heavy ion reactions is
reexamined using Boltzmann-Uehling-Uhlenbeck simulations.  We suggest that, depending on the incident energy and impact
parameter of the reaction, both entrance channel and exit channel effects can be at the origin of the observed behavior.
Specifically, fluctuations in the reaction mechanism induced by fluctuations in the collision rate,
as well as thermal bimodality directly linked to the nuclear liquid-gas phase transition are observed in our simulations.
Both phenomenologies were previously proposed in the literature, but presented as incompatible and contradictory interpretations of the experimental measurements.
These results indicate that heavy ion collisions at intermediate energies can be viewed as a powerful tool to study both
bifurcations induced by out-of-equilibrium critical phenomena, as well as finite size precursors of thermal phase transitions.
\end{abstract}


\maketitle

\section{Introduction}
The definition and characterization of phase transitions in finite systems is a fascinating multi-disciplinary topic that has been studied within different phenomenological applications since several decades \cite{fisher,davis}.
One interesting aspect is linked to the smoothing of phase transitions with finite number of constituents and the existence of specific scaling behaviors \cite{botet} signaling the proximity of a critical point.
Another specificity of finite systems is linked to the non-equivalence of statistical ensembles out of the thermodynamical limit \cite{gross}.
This implies that the phase structure of the system does not only depend on the Hamiltonian or evolution rule
but also on the procedure of data sampling.

Depending on the experimental condition and the data sorting criterium, different thermodynamical anomalies can therefore be observed and associated to the realization of the finite system counterpart of a thermodynamic phase transition \cite{gross}.
The concept itself of statistical ensemble has to be redefined, since the experimental conditions and the sorting technique can produce a virtually
infinite number of possible statistical ensembles \cite{Ellis,costeniuc,gorenstein}. Such extended ensembles can be coherently modelled by accounting for the experimental constraints including time-odd observables and collective flows \cite{gulmi_annphys,ison_flow}, and lead to predictions that can interpolate between the standard canonical, microcanonical and grandcanonical ensembles of macroscopic $(N,V,T)$ systems \cite{gaussian,Johal}.

For a temperature driven phase transition, in the particular case of a canonical (or close to canonical) sorting, the two-peaked or bimodal behavior of the order parameter distribution is known to be a robust observable of the phase transition \cite{zeroes,bimo,bimo_add,Pleimling,lehaut10}.
This observation has raised some interest in the heavy ion community, because it opens up the possibility of experimentally pinning down the nuclear Liquid-Gas phase transition through the experimental measurement of the size distribution of the heaviest cluster produced in properly sorted multifragmentation reactions \cite{bimodal_fg,bimodal_gargi}.  Inspired by these theoretical works, clear bimodality signals were evidenced in different sets of nuclear multifragmentation data \cite{bimodal_asy,bimodal_zmax}.

The first observations \cite{bimodal_asy} concerned the distribution of the normalized charge asymmetry between the heaviest $Z_{max}$ and the second heaviest $Z_2$  fragment, $a_2=(Z_{max}-Z_2)/(Z_{max}+Z_2)$  in events corresponding to a given centrality, selected in bins of light particle transverse energy $E_{\perp 12}$. From the experimental point of view, the observation appears very robust and appears in virtually all different multifragmentation experiments.
However, this observable is only loosely correlated to the order parameter \cite{botet,bimodal_fg}, and it is not proved that the sorting can be assimilated to a canonical one.  Moreover this very same signal was successfully reproduced by Quantum Molecular Dynamics (BQMD) \cite{zbiri,lefevre} and Boltzmann-Uehling-Uhlenbeck (BUU) \cite{mallik16} calculations where a memory of the entrance channel is clearly present and thermal equilibrium is not achieved.
The signal was interpreted in these studies as a dynamical bifurcation \cite{lefevre} of reaction mechanism, induced by fluctuations of the collision rate which leads to fluctuations of the collective momentum distribution as expected in complex non-linear dynamical systems.

Other successive experimental studies \cite{bimodal_zmax} concerned the asymptotic heaviest cluster charge $Z_{max}$, which is strongly correlated to the theoretical order parameter, namely the size of the heaviest cluster at the fragmentation time. In those studies an explicit canonical sorting was applied, based on the event-by-event measurement of the excitation energy. Also in that case a  bimodality was observed,  which would rather point towards a thermal phase transition. However, the data analysis is less direct and the imperfect estimation of the calorimetric excitation energy might deform the signal.

Therefore, the origin of the experimentally observed bimodality is still not clear.

In the previous dynamical approaches used to study the bimodality phenomenon \cite{zbiri,lefevre,mallik16}, the collision final state was  determined by the semiclassical one-body transport equation itself, considering simulations evolving until asymptotic times. However, these approaches  lack the necessary correlations to properly treat fragment formation in the exit channel, even if they are known to   very well describe the entrance channel of heavy-ion reactions at intermediate energy. For this reason, to have a quantitative reproduction of experimental data, the secondary decay of the dynamically formed primary fragments is typically treated in two-steps calculations, coupling the transport dynamics to a statistical model (or "afterburner") \cite{chomaz}. In this work, we follow this standard procedure \cite{gargi13} using a recently developed version of the Boltzmann-Uehling-Uhlenbeck (BUU) transport model \cite{gargi15} for the dynamical phase, and the Canonical Thermodynamical Model (CTM) \cite{dasgupta} for the de-excitation phase. Both models have been already successfully confronted to a large set of experimental data (see \cite{gargi15,dasgupta}).

We observe that, depending on the incident energy and impact parameter of the reaction, both bimodality mechanisms can appear, meaning that the
different scenario proposed in the literature are both potentially observable in heavy ion data.

Specifically, fluctuations in the stopping dynamics in central collisions lead to different reaction mechanisms that can coexist in the the sample characterized by a well defined value of the impact parameter. This gives rise to a bimodal behavior of the $Z_{max}$ distribution that can survive to the secondary de-excitation if the deposited energy is low enough, which happens at  incident energies in the Fermi energy domain (40 A.MeV).
At higher incident energies (100 A.MeV), focusing on binary mid-peripheral reactions, the fluctuations in the energy deposition leads to an excitation energy distribution for the quasi-spectator source which is close to the LG phase transition range. For these events, local equilibrium is achieved and a thermal bimodality is observed in agreement  with statistical expectations.

These results have important implications on the possibility of connecting bimodality signals to a
possible finite size precursor of phase coexistence. Specifically, inclusive measurements as in refs.\cite{bimodal_asy},
where the different reaction mechanisms are summed up with the only requirement of a quasi-complete
detection in the forward hemisphere, are more likely to be connected to the dynamical bimodality
observed in the QMD calculations. Conversely, if a careful selection of the binary character of the
reaction mechanism is performed as in refs.\cite{bimodal_zmax}, we can expect that
the signal might be ascribed to a thermal
effect, that is to the convex entropy intruder indicating the first order phase transition \cite{gross}.

For this first exploratory study, we concentrate on a single light symmetric system $^{40}Ca+^{40}Ca$. This does not allow yet to make quantitative comparisons with experimental data, which are left for future work. However, it was already observed that size \cite{lefevre} and Coulomb \cite{gulminelli03,mallik15,gargi09} effects might distort but not qualitatively compromise the bimodality signals. This means that the results of the present paper are expected to represent general trends which can be observed also with other heavier systems.

The paper is organized as follows. The models employed are briefly reviewed in Section \ref{sec:models}, and the coupling conditions between the dynamical and statistical treatment is detailed in Section \ref{sec:FO}. Our results concerning the different conditions of occurrence of the bimodality signals are given in Section \ref{sec:results}, and Section \ref{sec:concl} summarizes the paper.

\section{Dynamical and statistical models}\label{sec:models}
The BUU transport model calculation \cite{Mallik10,Dasgupta_BUU1} for heavy ion collisions starts with two nuclei in their respective ground states approaching each other with specified velocities and impact parameters. Calculations are done in a 200$\times 200\times 200 fm^3$ box. At t=0 fm/c the projectile and target nuclei are centered at (100 fm,100 fm,90 fm) and (100 fm,100 fm,110 fm). The ground state energies and densities of the projectile (mass number $A_p$) and target (mass number $A_t$) nuclei are constructed using the Thomas-Fermi approximation \cite{Lee,Mallik_thesis}. The Thomas-Fermi phase space distribution is then sampled using  Monte-Carlo technique by choosing test particles (we use $N_{test}=100$ for each nucleon) with appropriate positions and momenta.\\
\indent
As the the projectile and target nuclei propagate in time, the test particles move in a mean field and occasionally suffer two-body collisions , with probability determined by the nucleon-nucleon scattering cross section, provided the final state of the collision is not blocked by the Pauli principle. To explain clustering in multifragmentation, one needs an event by event computation in transport calculation. To do that, we have followed the recently developed computationally efficient prescription described in Ref. \cite{Mallik10} which lies midway between the original BUU calculation \cite{Dasgupta_BUU1} and the original fluctuation added model \cite{Bauer}. According to this prescription, the nucleon-nucleon collisions are computed  at each time step with the physical cross-section $\sigma_{nn}$ only among the $A_p+A_t$ test-particles belonging to the same event. For each event, if a collision between two test particles $i$ and $j$   is allowed, the method proposed in ref. \cite{Bauer} is  followed: the $N_{test}-1$ test particles closest to $i$ are  picked and the same momentum change $\Delta \vec{p}$ as ascribed to $i$ is  given to all of them. Similarly the $N_{test}-1$ test particles closest to $j$ are  selected and these  are  ascribed the same momentum change $-\Delta \vec{p}$ suffered by $j$. As a function of time this is continued till the event is over. For the mean-field propagation the Vlasov technique is employed: all test particles are used and the Lattice Hamiltonian method \cite{Lenk,gargi13} is used for calculating the mean field potential. This procedure is repeated for as many events as one needs to build up enough statistics. The details of BUU transport model calculation can be found in \cite{mallik16,Mallik10,Mallik_thesis}\\
\indent
At the end of the transport calculation at freeze-out stage, we get different clusters of finite number of test particles with known position and momenta. By knowing the number of test particles present in the cluster one can get the mass, and by knowing the position and momenta of these test particles one can calculate the potential and kinetic energies respectively. By adding kinetic and potential energy the excited state energy of the cluster can be obtained. However, to know excitation one needs to calculate the ground state state energy also. This is done by applying the Thomas Fermi method for a spherical (ground state) nucleus having mass equal to the cluster mass. Knowing PLF mass and its excitation, the freeze-out temperature is calculated by using the canonical thermodynamic model CTM\cite{dasgupta}.\\
Indeed CTM  can be used to calculate the average excitation per nucleon for a given temperature and mass number, and the relation is inversed to get the temperature from the output of the dynamical stage\cite{Mallik_thesis}.   In CTM, it is assumed that a system with $A_0$ nucleons at temperature $T$, has expanded to a higher than normal volume where the partitioning into different composites can be calculated according to the rules of equilibrium statistical mechanics. According to this model, the average number of composites with $a$ nucleons can be calculated from,
\begin{equation}
<n_{a}>=\frac{\omega_{a}Q_{A_0-a}}{Q_{A_0}}
\end{equation}
where, $\omega_{a}$ is the partition function of one composite with $a$ nucleons and $Q_{A_0}$ is the total partition function which can be calculated from the recursion relation,
\begin{equation}
Q_{A_0}=\frac{1}{A_0}\sum_{a}a\omega_{a}Q_{A_0-a}
\end{equation}
It is important to stress that eq.(2) is an exact expression for the canonical partition sum of the statistical model \cite{dasgupta}. Therefore, the only model dependence of eq.(1) arises from the chosen expression for the cluster energies and entropies. Though theses quantities naturally influence the cluster distribution, the qualitative observation of a phase transition around $T\approx 5$ MeV with a bimodal order parameter distribution, is a generic feature of the cluster model \cite{bimo_add,dasgupta,gargi09}. The details of the statistical model can be found in Ref. \cite{dasgupta}.

When the temperature is lower than the temperature associated to the nuclear Liquid-Gas phase transition $T_t\approx 5$ MeV, the final fragment distribution is almost independent of the assumed freeze-out volume and very close to the one predicted by standard evaporation models\cite{Tsang,Charity,gargi11}.

\section{Determination of freeze-out}\label{sec:FO}

We first intend to identify the time when the target and projectile nuclei are completely separated and they have reached their freeze-out stage so that one can safely stop the dynamical calculation there and switch over to the statistical one. To do that we have studied the evolution of the largest and second largest cluster with time and also have studied the isotropy of the momentum distribution inside these fragments as a function of time. Indeed in the binary collisions we consider the largest and second largest cluster are always the residues of projectile and target.
The first signal can therefore help us to determine the time when the projectile and target are completely separated while the second one will point to the attainment of thermalization of these residues. To state it more precisely, the time when the size of the second largest cluster will reach maximum is actually the time when one can consider the target and the projectile to have completely crossed each other and there is no overlap between them. These can then fragment and one can apply the statistical models to take care of that part.\\
\indent
Fig. 1 displays the variation of the average size of the largest and the second largest cluster with time for four different impact parameters ranging from central to peripheral collisions at the projectile beam energy of 100 MeV/A. In each case we have started the time evolution from 50 fm/c and continued till 300 fm/c. In the beginning, there was just one system comprising of both projectile and the target and hence the size of the largest cluster is $A_{max}=80$
for our $A_p=40$ on $A_t=40$ reaction, while that of the second largest being obviously zero at this time.\\
\indent
The nature of variation for both the largest and the second largest is similar for all the impact parameters shown from $b=0$ fm to $ b= 9 $ fm. The size of the largest cluster decreases gradually with time as the system fragments as well as there is evaporation of light clusters and nucleons.  In case of b=0 fm (central), the rate of decrease is maximum, while for b=9 fm it is the least. This is because the size of the participant zone is maximal for central collisions,   which results in faster disintegration and hence smaller size of the largest fragment. The size of the second largest starts from zero, gradually increases as the target and projectile crosses  and reaches a maximum when they are completely separated and then again decreases because of secondary decay, and settles to  a final value.  The evolution of the largest and that of the second largest cluster is  pretty similar after the second largest cluster reaches its maximum and the evolution coincides for the most peripheral collisions.  This is only because we are dealing with a symmetric collision, and would change if we would consider an asymmetric entrance channel.\\
\begin{figure}[!h]
\includegraphics[width=10cm,keepaspectratio=true]{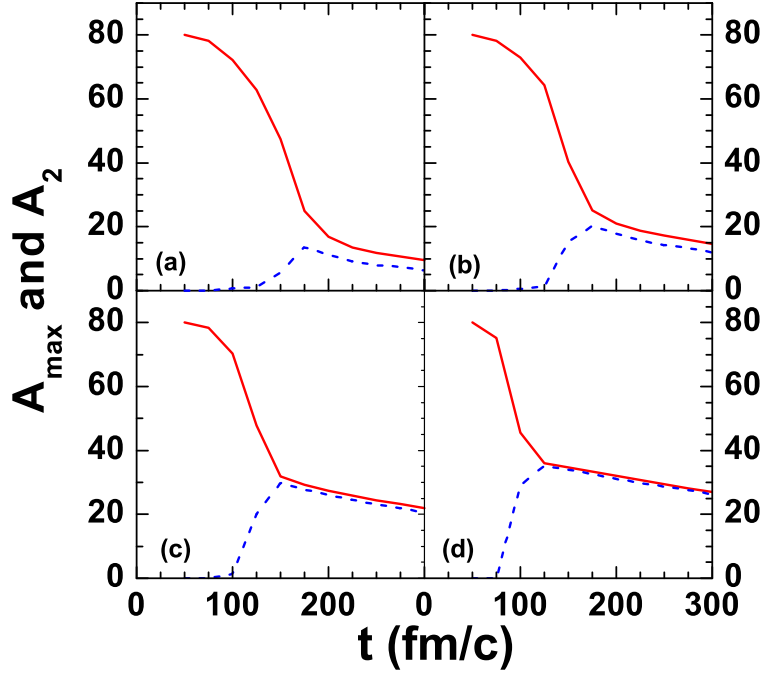}
\caption{(Color online)Variation of average mass of largest cluster $A_{max}$ (red solid lines) and second largest cluster $A_2$ (blue dashed lines) with time as calculated from BUU model for (a)$b=$0 fm, (b)$b=$3 fm, (c)$b=$6 fm, (d)$b=$9 fm at projectile beam energy 100 MeV/nucleon.}
\end{figure}
\indent
Fig. 2 displays the  time evolution of the average isotropy of the momentum distribution of the largest and second largest cluster. This observable indicates the thermalization of the final system, and hence the ideal time to stop the dynamical calculation. This is defined through the following equations. As described in the previous section, in BUU calculation, every nucleon of each event is represented by $N_{test}$ test particles. So the total number of test particle in every event is $(A_p+A_t)N_{test}$. Let for a given event, out of these $(A_p+A_t)N_{test}$ test particles, only N test particles form a cluster i.e. the mass of the cluster is $N/N_{test}$.\\ The average momentum of cluster along $k=$ $x$, $y$ and $z$ direction can be calculated from the relation
\begin{equation}
P_k=\frac{1}{N}\sum_{i=1}^{N} {p_{k_i}}
\end{equation}
where $p_{k_i}$ is the $k$ component of momentum of the $i$-th test particle.\\
The isotropy in momentum distribution can be defined as
\begin{equation}
I=\frac{\frac{1}{N}\sum_{i=1}^{N}(p_{x_i}-P_x)^2+\frac{1}{N}\sum_{i=1}^{N}(p_{y_i}-P_y)^2}{2\times\frac{1}{N}\sum_{i=1}^{N}(p_{z_i}-P_z)^2}
\end{equation}.
The quantity is defined such that it is less than 1 when the system is not fully thermalized and still there are some test particles  having significant momentum in the beam direction . This will reduce the isotropy and hence initially during the collision of the target and the projectile or during the crossing stage the isotropy is less than 1. With time it gradually increases and finally reaches the maximum possible value when full thermalization is achieved.
This is also almost the same time when the second largest cluster attained its maximum size as is shown in Fig 1. This freeze-out time varies from about  $t_{FO}=150$ fm/c at $b=9$ fm to about $t_{FO}=200$ fm/c at $b=0$ fm. For simplicity, we have
stopped  the dynamical calculation at $t=175$ fm/c for all impact parameters. Accounting for the precise impact parameter dependence of the freeze-out time would only marginally affect the distributions shown in this paper, and would not affect any of our conclusions which are essentially based on the qualitative properties of the distributions. A similar analysis was performed for the other case studied in this paper,  $E/A= 40$ MeV/A, and the coupling time was determined as approximately $t=400$ fm/c  in that case.
 \\
\indent
\begin{figure}[!h]
\includegraphics[width=10cm,keepaspectratio=true]{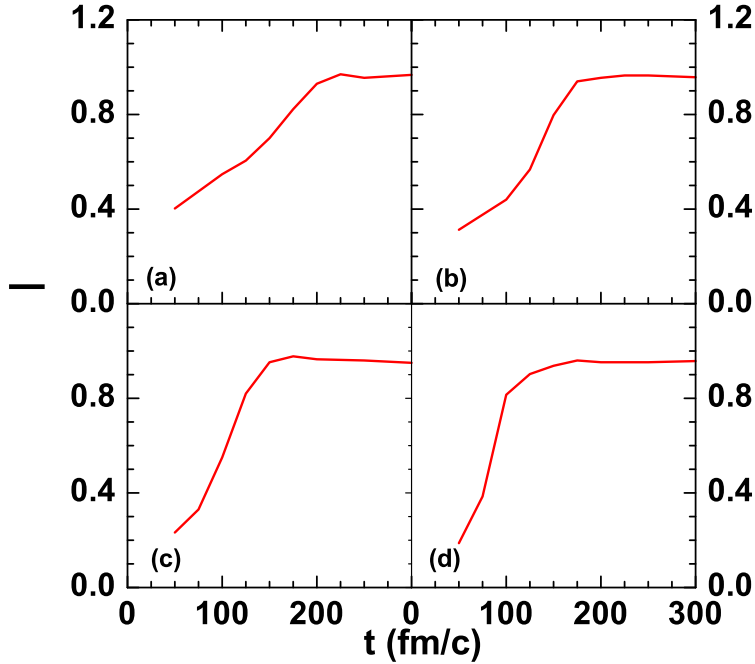}
\caption{(Color online)Variation of isotropy of momentum distribution ($I$) of the largest cluster with time calculated from BUU model for (a)$b=$0 fm, (b)$b=$3 fm, (c)$b=$6 fm, (d)$b=$9 fm at projectile beam energy 100 MeV/nucleon.}
\end{figure}

\section{Dynamical and statistical bimodality}\label{sec:results}

We now turn to study the behavior of the distribution of $A_{max}$, which represents the order parameter of the fragmentation transition\cite{botet,bimodal_fg}. We first examine the distribution obtained at the end of the dynamical calculation.

In Fig. 3 we have plotted the probability distribution of the average size of the largest cluster for four different impact parameters of varying centrality at $t$=175$fm/c$ where we have decided to stop the dynamical calculation. The distribution of asymmetry $a_2=(A_{max}-A_2)/(A_{max}+A_2)$, used in refs.\cite{bimodal_asy,zbiri,lefevre,mallik16} is also shown. For central collision($b$=0 $fm$), two peaks are seen in both distributions.

\begin{figure}[!h]
\includegraphics[width=10cm,keepaspectratio=true]{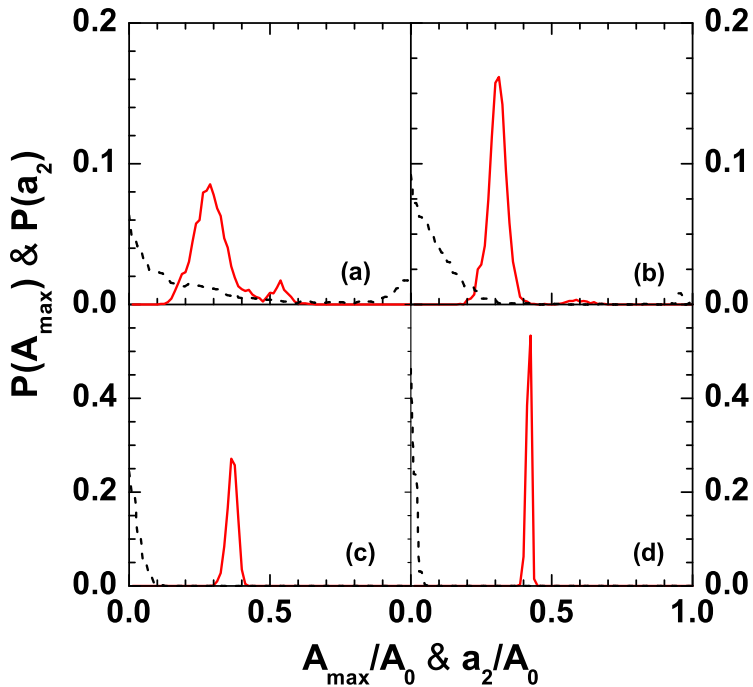}
\caption{(Color online)Probability distribution of largest cluster $P(A_{max})$ (red solid lines) and normalised mass asymmetry of two largest masses  $P(a_2)$ (black dashed lines) at constant projectile beam energy 100 MeV/nucleon but four different impact parameters (a)$b=$0 fm, (b)$b=$3 fm, (c)$b=$6 fm, (d)$b=$9 fm calculated from BUU model at freeze-out time $t$=175 $fm/c$. At each impact parameter 2000 events are simulated.}
\end{figure}
 We can interpret this observation as a dynamical bimodality very similar to the phenomenon described in ref.\cite{zbiri,lefevre}.
Fluctuations in the collision rates lead to fluctuations in the momentum distribution, that is in the degree of stopping of the reaction.
This is shown in the right panel of Figure 4 which displays the z component of the momentum (defined in Eq. 1) of the largest cluster for central collisions,
for the two classes of events corresponding to different cuts in the largest cluster size. We have fixed a mass cut of $A_{cut}=37$ to distinguish the two event classes as it corresponds to the minimum between the two peaks in fig. 3a. We can see that those with $A_{max}<A_{cut}$ have momentum similar to that of the initial projectile in forward (PLFs) as well as in the reverse (TLFs) direction as expected for an incomplete stopping leading to a binary collision. The other class($A_{max}>A_{cut}$) have $P_z$ nearly equal to zero indicating that they can be identified as completely stopped events.
The same information is given by the left panel of Fig.4 which displays the angular distribution of the largest cluster for the same events.  It is clearly seen from the figure that in events with $A_{max}>A_{cut}$, the fragments are emitted isotropically and hence they correspond to stopped events. On the other hand in events with $A_{max}<A_{cut}$, they are scattered either in the forward direction (PLFs) or in the backward direction (TLFs) clearly indicating their source to be  crossed events.\\
With increasing impact parameter, the peak at higher mass and higher asymmetry disappears implying negligible stopping while that at the lower mass shifts to the right because as the impact parameter increases, the participant zone decreases resulting in increase of the PLF/TLF size. The distribution also becomes sharper with increase of impact parameter  for the same reason.\\

\begin{figure}[!h]
\includegraphics[width=10cm,keepaspectratio=true]{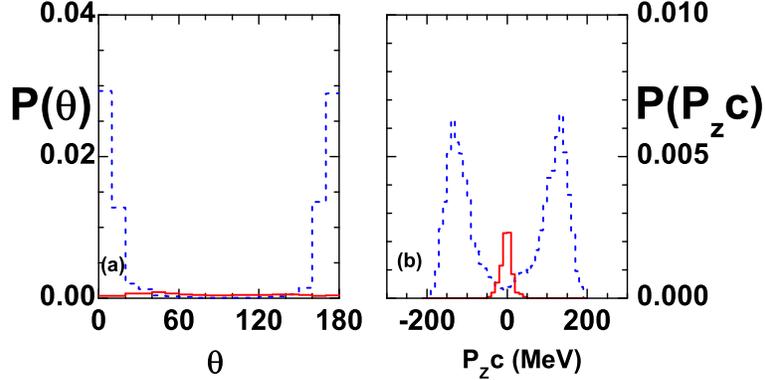}
\caption{(Color online)Largest cluster scattering angle (left panel) and momentum (right panel) probability distribution for $A_{max} \geq 37$ (red lines) and $A_{max}<37$ (blue lines) for central collisions ($b=$0 fm) at projectile beam energy 100 MeV/nucleon calculated from BUU model at freeze-out time $t$=175 $fm/c$. To study this, 2000 events are simulated. The average value of 10 degrees and 10 MeV are shown for angle and momentum respectively.}
\end{figure}
\indent

The distribution plotted in Fig.3 and 4 can be defined as freeze-out distribution and still evolve in subsequent time because of secondary decay.
If the excitation energy at the time of freeze-out is below the threshold of cluster emission, the secondary decay only involves particle emission.
In that case we expect $A_{max}$ to monotonically decrease in time, and the shape of the distributions to be  preserved.
This is also observed  if we simply run the calculation for a longer time. However, if the excitation energy of the heaviest cluster is higher, this latter can undergo multiple decay. This multifragmentation channel is not accessible in the transport equation because it needs the consistent inclusion of many-body correlations \cite{chomaz} which is out of the scope of transport models. Because of that, the fragment distributions obtained at asymptotic times in the transport calculation might not correspond to the asymptotic physical partition. To overcome this problem, the secondary decay is treated with the statistical CTM model, with inputs given event-by event by the transport calculation at the freeze-out time, as explained in section \ref{sec:models}.

\begin{figure}[!h]
\includegraphics[width=13cm,keepaspectratio=true]{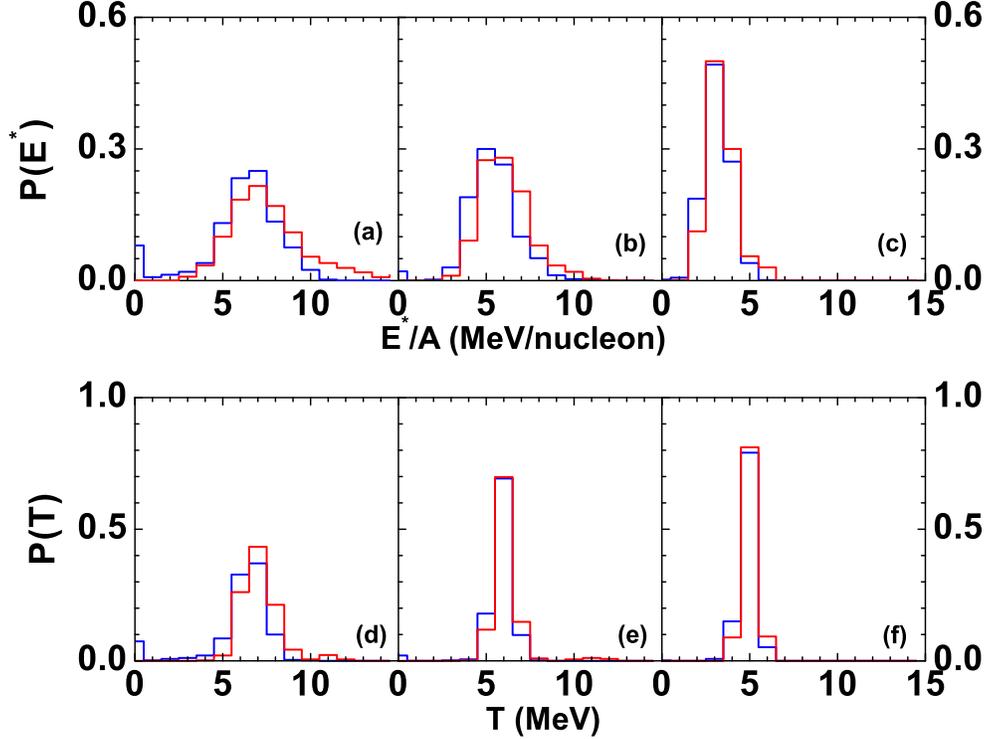}
\caption{(Color online)Excitation ($E^*$) (upper panels) and temperature ($T$) (lower panels) probability distribution for largest (red dash dotted lines) and second largest cluster (blue dotted lines) at constant projectile beam energy 100 MeV/nucleon but two different impact parameters $b=$0 fm (left panels), $b=$3 fm (middle panels) and $b=$6 fm (right panels). At each impact parameter 2000 events are simulated. The average value of 1 MeV/nucleon and 1 MeV are shown for excitation and temperature respectively.}
\end{figure}

Further details on the evaluation of temperature and excitation energy from the BUU calculation can be found in refs.\cite{gargi13,gargi15}.

In Fig. 5 we have plotted the distribution of excitation energy and temperature of the largest as well as the second largest cluster for three different impact parameters. The step size selected for displaying these distributions are 1 MeV/A for the excitation energy and 1 MeV for the temperature. For the excitation energy, the distribution is more or less similar in shape for both  impact parameters; there is a small peak at low excitation that corresponds to the second largest cluster of the stopped event which is small in size and has less excitation. Using these excitation energies from the transport code as input to the statistical model code, temperature is calculated and its distribution is plotted. The obtained temperaure distributions of the largest and second largest fragment show a good agreement, providing an extra test of equilibrium for the freeze-out configuration. At b=3 and 6 fm, the distribution is quite sharp strongly indicating its connection to the phase transition during which the temperature remains constant. This will be further established from the next figure.\\
\begin{figure}[!h]
\includegraphics[width=11cm,keepaspectratio=true]{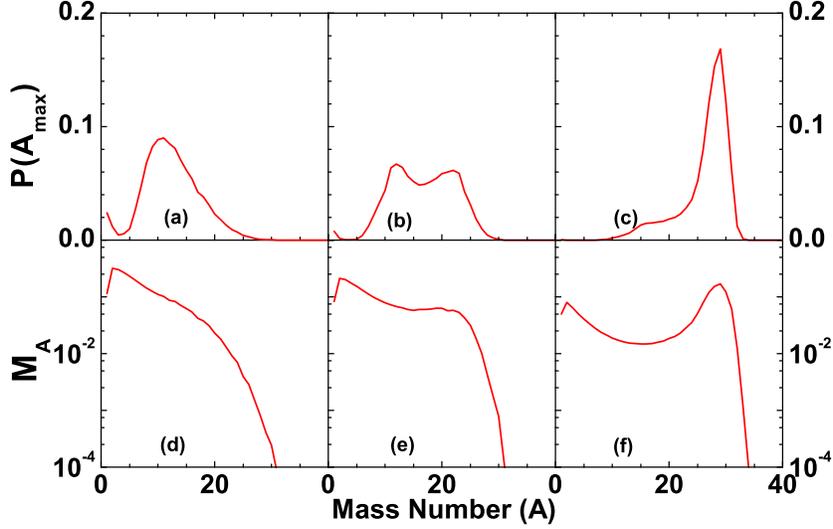}
\caption{(Color online) Final largest cluster probability distribution (upper panels) and mass distributions (lower panels) studied after CTM calculation at constant projectile beam energy 100 MeV/nucleon but three different impact parameters $b=$0 fm (left panels), $b=$3 fm (middle panels) and $b=$6 fm (right panels). At each impact parameter 2000 events are simulated.}
\end{figure}
\indent
In Fig. 6 we have plotted the probability distribution of the largest cluster as well as the total multiplicity  for these three impact parameters.  These distributions have been calculated after switching over to the statistical code from the transport one. The ones at b=0 fm are structuresless and typical of multifragmentation reactions: the average excitation energy is so high in that case that both fully stopped and incompletely stopped events undergo multiple decay. As a consequence, the bimodality signal observed in Fig.3 disappears.\\
\indent
At higher impact parameter, the situation is reversed. The probability distribution of the largest cluster now shows a bimodal behaviour  which is indicative of existence of two phases simultaneously. The mass distribution on the lower panel can be seen as a superposition of a multifragmentation distribution (predominant the case of b=0) with a residue distribution (clearly visible the case b=6). This illustrates the well-known fact  that in the case of heavy ion reactions the ordered phase can be associated to compound nucleus evaporation, while the disordered phase can be associated to multifragmentation.
\begin{figure}[!h]
\includegraphics[width=10cm,keepaspectratio=true]{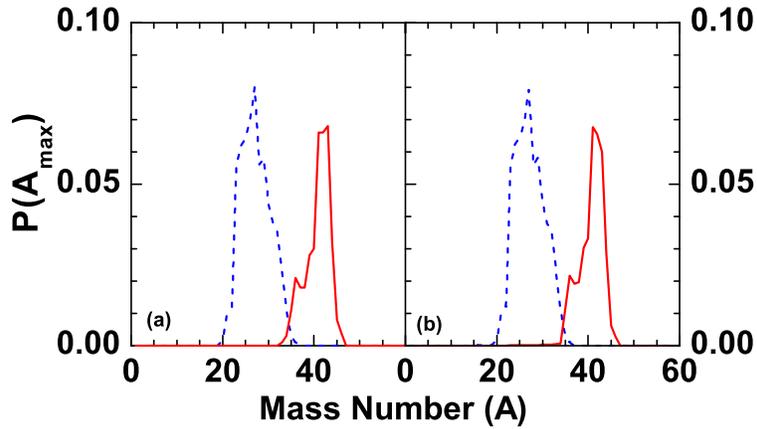}
\caption{(Color online)Largest cluster probability distribution for crossed events (blue lines) and stopped events (red lines) studied after BUU model calculation (left panel) and CTM calculation (right panel) for central collisions ($b=$0 fm) at projectile beam energy 40 MeV/nucleon. In order to study this 500 events are simulated and the BUU calculation is stopped at $t=$400 $fm/c$. }
\end{figure}
\indent
The dynamical bimodality displayed in Fig.3 does not survive secondary decay (and is therefore not detectable experimentally) because the excitation energy deposited in the quasi-projectile source is too high. This however strongly depends on the entrance channel conditions.
In particular we may expect that lower bombarding energy might lead to a situation where the freeze-out distribution is not distorted by secondary decay. This is shown in Fig. 7, where  we have plotted the largest cluster probability distribution for central collisions at 40 MeV/nucleon both after transport calculation, and after the statistical model calculation. We have plotted separately the crossed and the stopped events in both the cases. These events are separated following the prescription described in Fig 4 based on the momentum distribution. In both the cases the largest cluster probability distribution shows a dynamical bimodality, and the distribution both after transport calculation and that after decay is almost same indicating  a small contribution of secondary decay in this energy domain. For completeness, the asymmetry distribution before and after secondary decay is shown in Fig.8 for a representative impact parameter. As already observed in the case of the higher beam energy in Fig.3, the information of the $A_{max}$ distribution is perfectly consistent with the information given by the asymmetry. \\
\begin{figure}[!h]
\includegraphics[width=7cm,keepaspectratio=true]{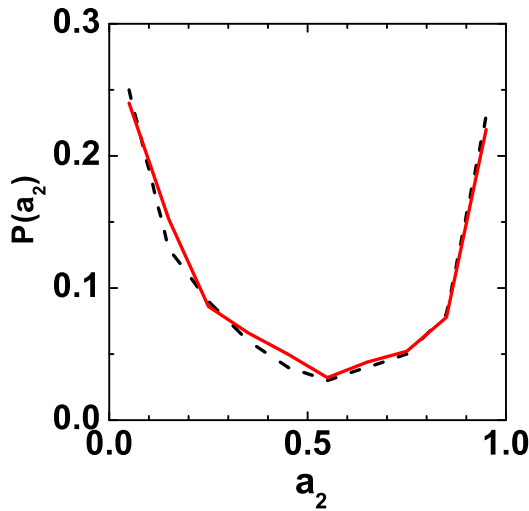}
\caption{(Color online) Probability distribution of normalised mass asymmetry of two largest masses  $P(a_2)$ studied after BUU model calculation (black dashed line) and CTM calculation (red solid line) for central collisions ($b=$0 fm) at projectile beam energy 40 MeV/nucleon. To study this 500 events are simulated and the BUU calculation is stopped at $t=$400 $fm/c$. }
\end{figure}
\indent
Another situation leading to a very small effect of secondary decay, and therefore a potential observation of the dynamical bimodality,  is the case of a peripheral spectator source at high incident energy, as we can also infer from the right panels of Fig.5. This is indeed the kinematic situation where a dynamical bimodality was observed in BQMD calculations \cite{lefevre}. In our calculations, the collision rate is not significant enough at $b=6$ and $E/A=100$ MeV/A to produce any stopped event, at variance with that calculation. This is probably due to the small size of  the $Ca+Ca$ system studied in this paper. Indeed the average collision rate, as well as its fluctuations, increases with increasing mass of the colliding system \cite{lopez}.
It is therefore probable that both dynamical and thermal bimodalities could be observed at the same time in the $Au+Au$ system, consistent with ref.\cite{lefevre}. This point is currently under study.

\section{Conclusions}\label{sec:concl}

In this paper we have analyzed the largest fragment size distributions for the $Ca+Ca$ system at two different bombarding energies,
as predicted by a two-step model. The entrance channel dynamics is described by the BUU transport equation, which is coupled to the statistical CTM decay model at the time of local equilibration of the primary fragments produced in the collision.\\
\indent
We have shown that different initial conditions can lead to the observation of bimodal distributions. When the collision rate is sufficiently high (as it is only the case in central collisions, for the light system analyzed), two different reaction mechanisms can coexist at the same impact parameter, corresponding to fully stopped and partially stopped events. This dynamical entrance channel effect leads to a bimodal distribution of the largest cluster, which can persist in the asymptotic stage if the excitation energy deposited in the non-stopped events is below the threshold of cluster emission. This is obtained if the incident energy is low enough (40 MeV/A in the present application). The same behavior can also be potentially obtained if the degree of transparency of the non-stopped events is sufficiently important, or in other terms if the fluctuation in the momentum distribution is very high. Such a behavior was reported in ref.\cite{lefevre} for the $Au+Au$ system.\\
\indent
When non-stopped binary events are selected, and the average excitation energy deposited in the quasi-projectile source is sufficiently large, as well as its fluctuation, the collision passes through a freeze-out stage that can be assimilated to a quasi-canonical ensemble at a temperature close to the LG transition temperature. This happens in our calculations for semi-central collisions at high bombarding energy. In that case, the system undergoes a thermal phase transition which signalled again by a bimodal distribution of the order parameter.\\
\indent
These results indicate that heavy-ion collisions can be used as a laboratory to study both types of bimodalities which have been proposed in the literature.
Indeed we have shown that the two bimodality mechanisms are associated
in the transport model to different time scales of the reaction, and to different energy regimes.
This means that both dynamical bifurcations and first order phase transition
could be potentially measurable with heavy ion collisions.

The reaction mechanism out-of-equilibrium bifurcation due to the non-linearity of the dynamical evolution can be studied  within a proper selection of the entrance channel, while the thermal bimodality signaling the liquid-gas phase transition requires an additional control on the reaction mechanism.
It is possible that the experimental studies reported in refs.\cite{bimodal_asy} correspond to the dynamical bimodality while the ones of refs.\cite{bimodal_zmax}, where a careful control of the reaction mechanism was performed, might correspond to the thermal phenomenon.\\
\indent
To confirm this hypothesis, further calculations in the same conditions as studied experimentally are in progress.

\end{document}